\documentclass[conference]{IEEEtran}
\IEEEoverridecommandlockouts

\usepackage{cite}
\usepackage{url}
\usepackage{amsmath,amssymb,amsfonts}
\usepackage{algorithmic}
\usepackage{graphicx}
\usepackage{textcomp}
\usepackage{xcolor}
\usepackage{comment}
\usepackage{authblk}

\def\BibTeX{{\rm B\kern-.05em{\sc i\kern-.025em b}\kern-.08em
    T\kern-.1667em\lower.7ex\hbox{E}\kern-.125emX}}

\begin{document}

\title{Wireless Spectrum in Rural Farmlands: Status, Challenges and Opportunities}

\author{
    Mukaram Shahid\IEEEauthorrefmark{1},
    Kunal Das\IEEEauthorrefmark{1},
    Taimoor Ul Islam\IEEEauthorrefmark{1},
    Christ Somiah\IEEEauthorrefmark{1},
    Daji Qiao\IEEEauthorrefmark{1},
    Arsalan Ahmad\IEEEauthorrefmark{1},\\
    Jimming Song\IEEEauthorrefmark{1},
    Zhengyuan Zhu\IEEEauthorrefmark{1},
    Sarath Babu\IEEEauthorrefmark{1},
    Yong Guan\IEEEauthorrefmark{1},
    Tusher Chakraborty\IEEEauthorrefmark{2},
    Suraj Jog\IEEEauthorrefmark{2},\\
    Ranveer Chandra\IEEEauthorrefmark{2},
    Hongwei Zhang\IEEEauthorrefmark{1}}
\affil{
\IEEEauthorrefmark{1}Iowa State University,

\IEEEauthorrefmark{2}Microsoft
}

\maketitle

\begin{abstract}
Due to factors such as 
low population density and expansive geographical distances, network deployment falls behind in rural regions, 
leading to a broadband divide. Wireless spectrum serves as the blood and flesh of wireless communications. Shared white spaces such as those in the TVWS and CBRS spectrum bands offer opportunities to expand connectivity, innovate, and provide affordable access to high-speed Internet in under-served areas without additional cost to expensive licensed spectrum. 
    However, the current methods to utilize these white spaces are inefficient due to very conservative models and spectrum policies, 
    causing under-utilization of valuable spectrum resources. This hampers the full potential of innovative wireless technologies that could benefit farmers, small Internet Service Providers (ISPs) or Mobile Network Operators (MNOs) operating in rural regions. 
This study explores the challenges faced by farmers and service providers when using shared spectrum bands to deploy their networks while ensuring maximum system performance and minimizing interference with other users. 
    Additionally, we discuss how spatiotemporal spectrum 
    models, in conjunction with database-driven spectrum-sharing solutions, can enhance the allocation and management of spectrum resources, ultimately improving the efficiency and reliability of wireless networks operating in shared spectrum bands.
\end{abstract}


\section{Introduction}

Rural America is a major element in the fabric of the United States, contributing to its overall cultural, social, and economic landscape. Like water and electricity, broadband access has become an essential utility. Despite having a critical role in sustaining the nation's economy, the rural farmlands of the United States face a serious challenge: the broadband 
divide \cite{Rural_Broadband}. To illustrate, much like the impact of the Rural Electrification Act of 1930 \cite{Rural_Electrification},
which brought electricity to farms and hence changed the overall farming landscape, providing Internet access to these farms is of equal (if not more) significance. Many unique and important functions of the farming community, for instance, extended-reality (XR) based tele-operation of unmanned aerial and ground vehicles in precision agriculture require high-speed Internet connectivity.

However, expanding broadband coverage in rural regions faces significant economic challenges. According to a recent report from the Federal Communications Commission (FCC), nearly 22.4\% of Americans residing in rural regions and 27.7\% of those in tribal lands lack access to 
fixed terrestrial broadband meeting the minimum download/upload capacity requirement of 25/3 Mbps per user, which is in stark contrast to the fact that only 1.5\% of the urban population lack broadband access \cite{fcc2020}. In addition, the current market model lacks incentives for rapid, comprehensive rural broadband expansion. For instance, the cost of expanding broadband infrastructure in remote areas doubles that in urban centers, while the expected revenue tends to be ten times lower. This demotivates mobile network operators (MNOs) and Internet Service Providers (ISPs) from investing in rural regions.

To help mitigate rural broadband cost, license-free or shared access to spectrum can help incentivize MNOs and ISPs in their rural expansion, and it can also enable farmers to deploy their own private IoT and broadband networks across farms. However, the current spectrum management paradigm of 
Federal Communications Commission (FCC) and the National Telecommunications and Information Administration (NTIA) is such that they have allocated most of the public spectrum on license-based access where only the licensee has the authority to transmit in those spectrum bands while conforming to the commission's rule. This license-based spectrum access ensures that the Primary User (PU) or licensee enjoys protection from harmful interference, safeguarding the integrity of their deployed networks. But licensees often don't fully utilize the spectrum, leaving portions unused across various spatial and temporal scales. These unused spectrum, commonly termed ``White Spaces", present opportunities for Secondary Users (SUs) (i.e., non-licensee users) to access them while minimizing potential interference to the PUs. 
In the meantime, as spectrum demands rapidly increase over time, it becomes more and more difficult to completely vacate the spectrum bands already allocated to existing users and re-allocate them to new users. 
Therefore, the white-space-opportunity and the difficulty of complete spectrum re-allocation calls for a new spectrum era where the spectrum can be managed as a common pool resource and dynamically allocated and shared 
based on in-situ demand by individual users. 

In rural America, dynamic spectrum sharing in the low bands and mid-bands such as 
the TVWS (Television White Spaces) and CBRS (Citizens Broadband Radio Service) Bands are particularly promising, thanks to their favorable propagation characteristics for potential large coverage.
\cite{DB_Critique},\cite{DB_Implementation}. 
    However, deploying high-power wide-area networks on shared spectrum bands poses unique challenges, especially when it comes to ensuring specific Quality of Service (QoS) for critical applications. Existing spectrum sharing and management frameworks overlook these challenges by allocating spectrum in a very conservative manner, resulting in low spectrum use efficiency and inadequate user experience. Furthermore, these frameworks prioritize protecting PUs, leaving SUs vulnerable to interference without any guarantee of protection. Accordingly, MNOs tend to hesitate to invest in deploying their networks in
    shared spectrum bands. 

To address the rural broadband challenge and to unleash the full potential of dynamic spectrum sharing in rural America, this article explores the status of spectrum utilization in rural regions, the challenges faced by operators and small ISPs, and the opportunities available to solve these challenges. Section II introduces the ARA \cite{ARA:vision} wireless living lab, the NSF PAWR platform on rural broadband. Section III examines the current spectrum utilization in rural regions, focusing on key bands from the National Spectrum Strategy \cite{national_spectrum_strategy}. Section IV examines the implications of spectrum-sharing rulings by FCC and NTIA. Section V proposes a Physics-Informed Neural Network architecture for modeling rural wireless channels, and evaluate it using real-world measurement data from the ARA PAWR testbed. Section VI investigates open research questions revolving around physics-informed modeling and protocol design, as well as the benefits of spectrum sensing-based spatiotemporal modeling in dynamic spectrum management. The last section shares concluding remarks.

\section{ARA - A Rural Radio Dynamic Zone for Spectrum Research}

Rural communities are sparsely distibuted with acres of farmlands surrounding them. These areas have unique network requirements that differ from the urban counterpart. For instance, the spectrum demand in rural areas varies depending on seasons as well as the agriculture activities such as planting and harvesting. Traditional broadband technologies, such as fiber, do not considered as a viable solution for the last mile applications in rural regions, thereby necessitating innovative solutions to meet the requirements of rural broadband. 

\begin{figure}[!htbp]
        \centering
        \includegraphics[width=\columnwidth]{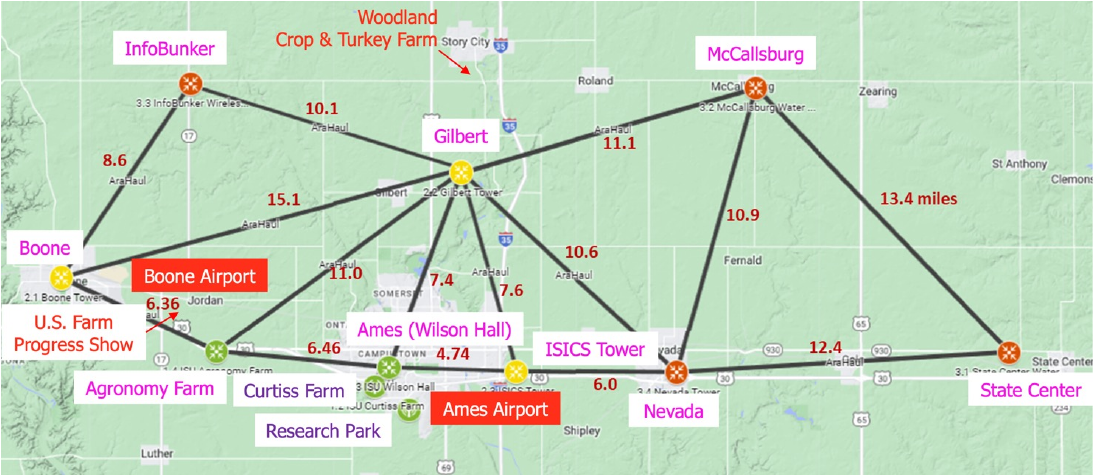}
        \caption{The ARA Wireless Living Lab in Central Iowa}
        \label{fig: ARA Deployment}
\end{figure}

To enable rural-focussed wireless innovation, we deploy 
ARA Wireless Living Lab~\cite{ARA:vision}, which is the fourth and final large-scale experimental platform under the NSF Platform for Advanced Wireless Research (PAWR) program. ARA spans an area over 60~km in diameter across the Iowa State University~(ISU) campus surrounding research and producer farms as well as rural communities of Central Iowa. Four wireless Base Stations~(BSes)---Agronomy Farm, Curtiss Farm, Research Park, and Wilson Hall (shown as green dots in Fig.~\ref{fig: ARA Deployment})---and 20+ User Equipment~(UE) have been deployed and made available to the community for advanced wireless and application research. Three additional BSes (shown as yellow dots in Fig.~\ref{fig: ARA Deployment}) and up to 30 additional UEs are planned to be deployed and made available by the end of summer 2024.

With the coverage of such a large and strategic rural geographical area, ARA serves as a Radio Dynamic Zone~(RDZ) where multiple users and services can dynamically share the spectrum across space, time, and frequency domains \cite{Radio_Dynamic_Zones}. These zones, spanning tens or hundreds of square kilometers, are designed to test the viability of spectrum-sharing solutions in real-world deployments. ARA provides a robust, multi-spectral infrastructure that ranges from TV White Spaces (TVWS) to millimeter-wave (mmWave) frequency bands, covering an area of 155 square kilometers. Such an extensive footprint makes ARA an ideal platform for implementing and evaluating dynamic spectrum sharing solutions in rural regions. In addition, the RF-silent environment around ARA enables testing of spectrum sharing algorithms in noise- and interference-free environments.

\section{Exploring the Present Spectrum Utilization}
\label{sec:exploring_spectrum}

Spectrum rights are often referred as the new rights on the water resource~\cite{Spectrum_Rights_as_Water_Rights}. As a critical resource, wireless spectrum is an inevitable component in sectors sensitive to national security.
The increasing demand for bandwidth highlights the limitations of traditional licensed or exclusive spectrum access. Current policies distribute spectrum rights uniformly across the country. However, such a one-size-fits-all policy often leads to under-utilization of the spectrum. For example, the 2900--3100~MHz band is vital for Maritime Navigational Systems (MNS) near the coastal areas and remains underutilized in landlocked regions, necessitating the need for a finer country-level policy formulation to better address the needs of area-specific use-cases~\cite{Polycentric_Management}. For instance, area-specific allocation enables the use of spectrum for agricultural  applications such as plant monitoring and phenotyping.
Adopting a dynamic spectrum allocation policy could unlock new opportunities  and ensure that the spectrum is allocated to the right applications.

\begin{table*}[htb]
    \centering
    \caption{Wireless Spectrum Usage near ARA's Agronomy Farm Base Station}
    \label{tab:reduced_table}
    \footnotesize 
    \setlength{\tabcolsep}{4pt} 
    \renewcommand{\arraystretch}{1.2} 
    \begin{tabular}{||l|l|p{1.8cm}|p{2.4cm}|p{2cm}||} 
        \hline
        \centering \textbf{Spectrum Band} & \centering \textbf{Primary User} & \textbf{Span (MHz)} & \textbf{Avg. Occupancy} & \textbf{95\textsuperscript{th} Percentile} \\ [0.3ex] 
        \hline\hline
        TVWS & TV Broadcasters & 470--698 & 46.8\%& 55.6\%\\ 
        \hline
        AMT & Aeronautical Mobile Telemetry Downlink & 1435--1525 & 0.85\% & 3.97\% \\
        \hline
        
        MNS & Maritime Navigation System & 2900--3100 & 0\% & 0.01\%  \\
        \hline
        
        ASS / EESS & Earth Exploratory Satellite Service and Amature Radio & 3100--3450 & 0.2\% & 0.678\%  \\
        \hline
        CBRS & Citizens Broadband Radio Service  & 3550--3700 & 0.678\% & 2.16\% \\
        \hline
        Federal FS / MS & Federal Fixed Service and Mobile Service & 4400--4940 &0.17\%&3.1\% \\ 
        \hline
        UAS & Non-Payload Unmanned Aircraft Systems & 5030--5091& 0.8\% &2\%  \\ 
      
        \hline
 
    \end{tabular}
\end{table*}

To enable effective spectrum sharing in any given area, it is essential to understand the availability of the wireless spectrum as well as its spatial and temporal utilization. Moreover, it is important to recognize the limitations of existing spectrum-sharing mechanisms. Using the Keysight RF sensors, we conducted a measurement study on various spectrum bands of national interest, as specified in the National Spectrum Strategy~\cite{NationalSpectrumStrategy}, as well as the other federal bands. The study was conducted at the Agronomy Farm site of the ARA wireless living lab. The results from our measurement study, summarized in TABLE~\ref{tab:reduced_table}, clearly indicate that a significant range of wireless spectrum is available near the Agronomy Farm BS site. In fact, the TVWS band tops in spectrum occupancy, with 47\% of channels are being occupied, whereas the frequency bands such as the AMT band (1435--1525 MHz), MNS band (2900--3100 MHz), and ASS/EESS band (3100--3450 MHz) are rarely used during the time of our measurement. The CBRS band is also highly underutilized due to the very sparse CBRS deployment in Central Iowa.

\subsection{{{Temporal Dynamics and Spatial Diversity}}}

\begin{figure}[!htbp]
        \centering
        \includegraphics[width=1\columnwidth]{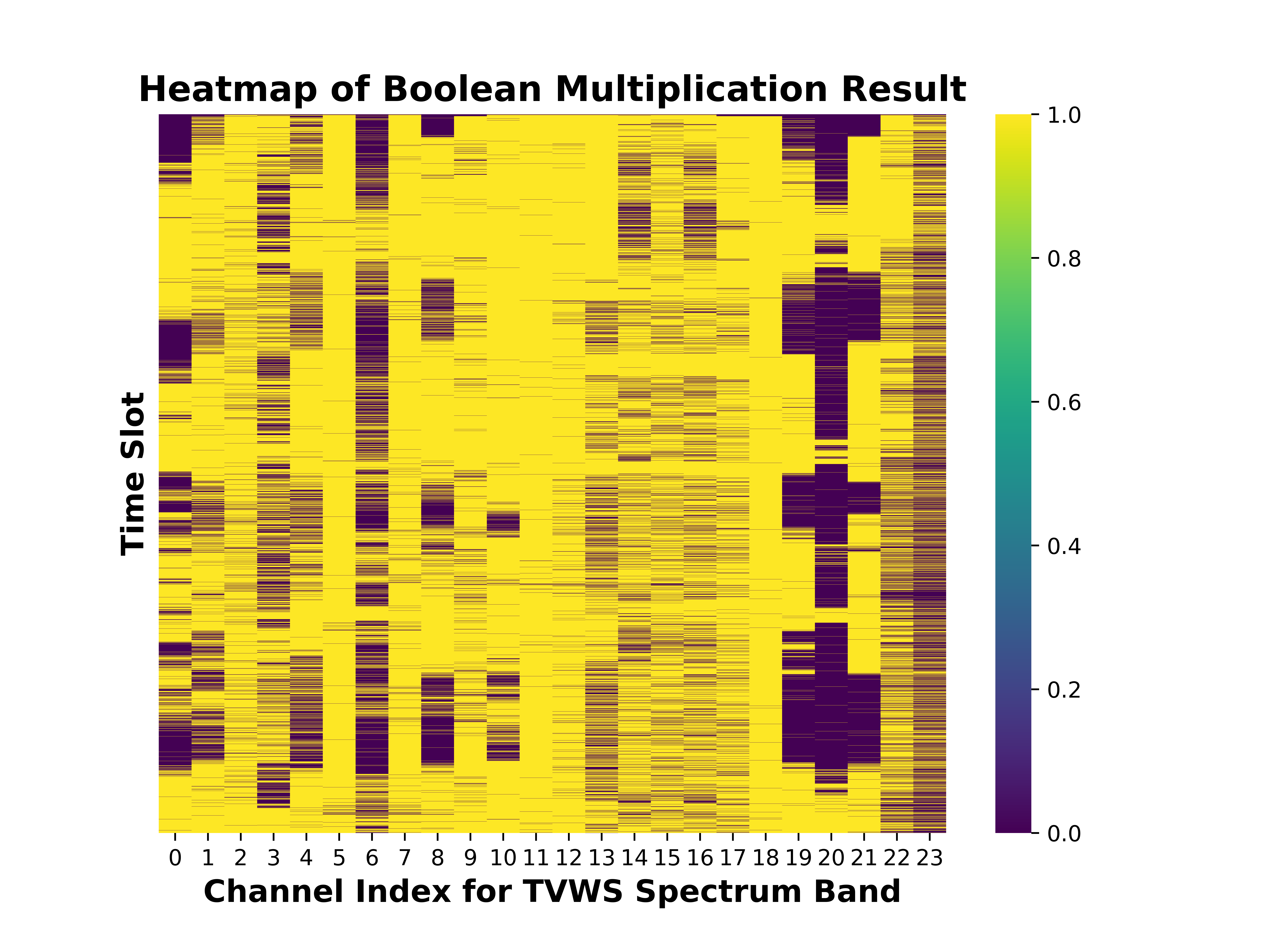}
        \caption{Spatiotemporal Diversity in the TVWS Band}
        \label{fig: Spatial Diversity}
\end{figure}

To analyze the spatial and temporal diversity, additional measurement studies were carried out specifically on the TVWS band and the 700 MHz band licensed to the Mobile Network Operators~(MNOs). We collected the spectrum usage data near the Wilson Hall and Agronomy Farm base station sites, which are 6 miles apart, to understand the correlation between spectrum usage at a suburban area and at a farm location. In Fig.~\ref{fig: Spatial Diversity}, the x-axis denotes the indices of broadcast TV channels spanning 470--608 MHz with each channel being 6~MHz wide,  and the y-axis represents the temporal occupancy of these channels for a period of 15~minutes, where each time slot is 1~second long. Yellow slots denote the spectrum availability in at least one of the two sites considered. On the other hand, the dark slots indicate the spectrum occupancy at both sites. In addition to the temporal dynamics, which can be clearly observed from Fig.~\ref{fig: Spatial Diversity}, there exists a unique spatial diversity in the occupied spectrum band influenced by terrain,  vegetation, weather, and building structures, impacting the transmitter link budget and, thereby resulting in significant differences in the occupied spectrum.

\subsection{{Dynamic Spectrum Capacity}}

Fig.~\ref{fig: Spatial Diversity} contradicts the common belief that the overall spectrum capacity remains constant over time, i.e., it illustrates the dynamic behavior of spectrum occupancy across bands. The occupancy continuously evolves not only over space, due to the random distribution of RF transmitters, but over time as well. The spectrum bands are not utilized consistently, leaving ample white spaces which could be effectively utilized with a suitable spectrum management solution. The unused white spaces offer opportunities for farmers intending to deploy their own private 5G networks for application-specific use cases.

\subsection{{Fragmented Spectrum}}

Spectrum fragmentation presents challenges in the management and sharing of the spectrum, specifically when multiple users with diverse bandwidth requirements use it for their applications. The fragmentation limits the users with wide bandwidth requirements and increases the risk of unwanted interference. While the MNOs utilize carrier aggregation in 5G stand-alone deployments with mid-band and high-band spectrum solutions leveraging sophisticated equipment, small ISPs and farmers deploying networks with Software Defined Radios~(SDRs) or less advanced equipment require better coordination techniques along with improved carrier aggregation and spectrum management to minimize out-of-band spectrum leakage.

\subsection{{Under-Utilization of Licensed Spectrum Bands}}

\begin{figure}[!htbp]
        \centering
        \includegraphics[width=1\columnwidth]{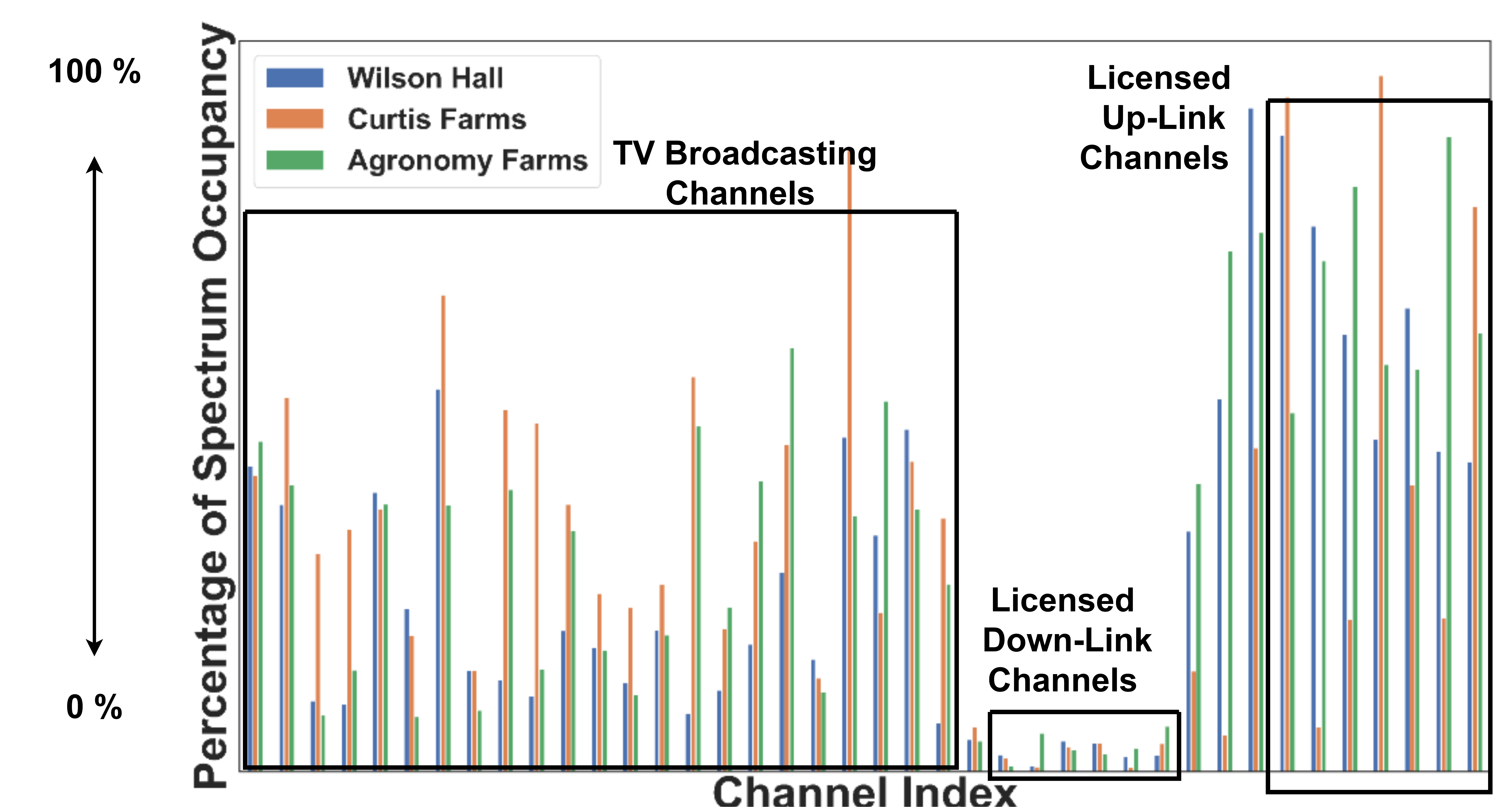}
        \caption{Under-utilized Licensed Spectrum}
        \label{fig: Under Utilized Spectrum}
\end{figure}

Fig.~\ref{fig: Under Utilized Spectrum} shows the spectrum occupancy of 470--746~MHz band at three ARA base station sites---Wilson Hall, Curtiss Farm, and Agronomy Farm. Channels 0--22 form the unlicensed TV broadcast spectrum bands, whereas the FCC licenses Channels 23--39. It is interesting to observe that among the spectrum that was licensed to MNOs for their broadband operations, the up-link channels are heavily utilized, however, the down-link channels are under-utilized during the time of our measurement. Such an under-utilization of the licensed spectrum could be subleased or used on a sharing basis for small operators to enable rural broadband.

From the ground measurements shown in TABLE~\ref{tab:reduced_table} and Fig.~\ref{fig: Under Utilized Spectrum}, it can be observed that many spectrum bands allocated to the federal departments are not being used all the time, indicating the importance of effective spectrum management solutions for driving innovative co-existence of multiple users while ensuring national security. As highlighted in the National Spectrum Strategy~\cite{NationalSpectrumStrategy}, the government is taking decisive action in re-purposing some spectrum bands studied in the above measurement while finding new solutions for spectrum sharing and co-existence.

\section{{Limitations of the current Spectrum Sharing Mechanisms Used by the FCC}}

\begin{figure}[!htbp]
        \centering
        \includegraphics[width=1\columnwidth]{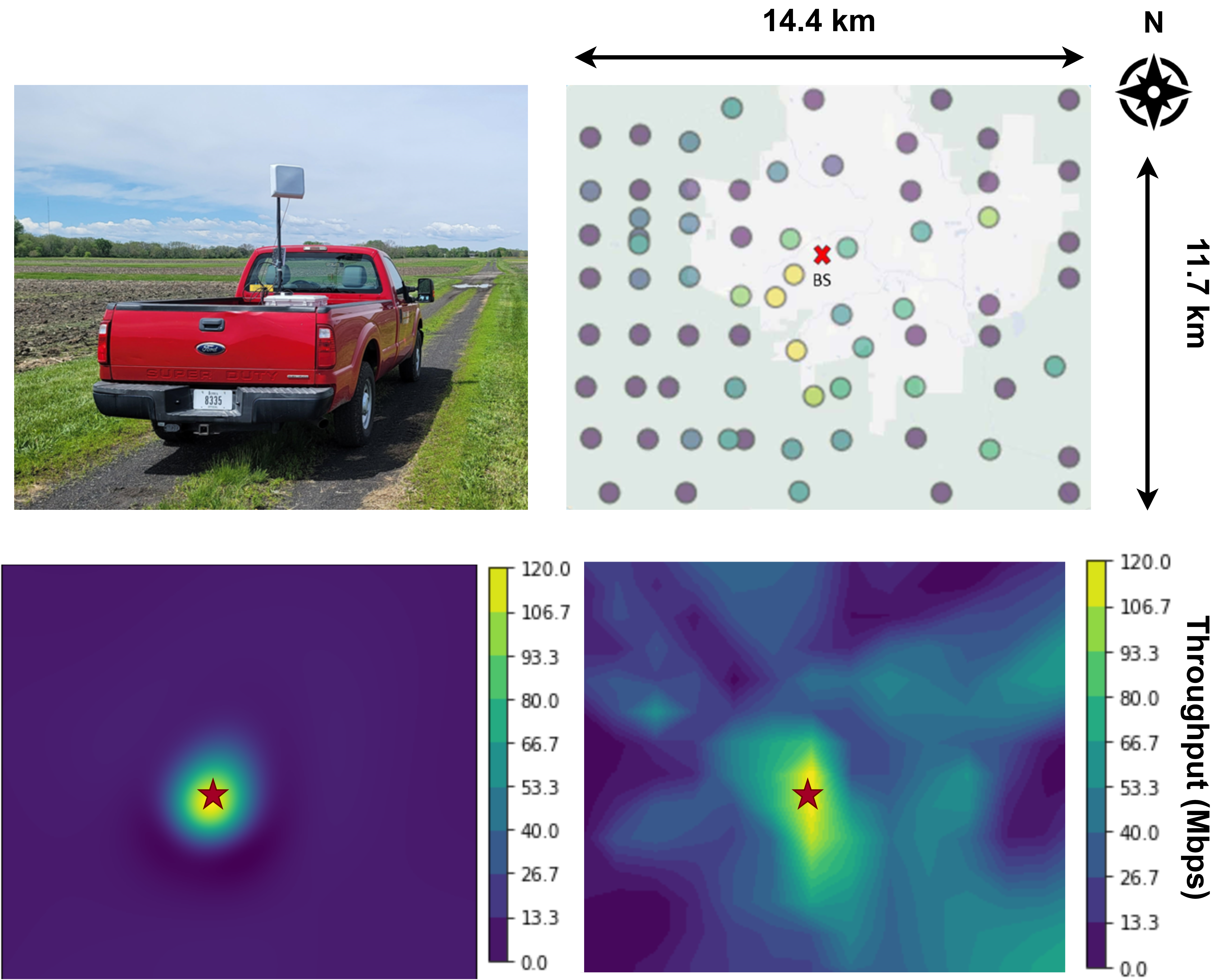}
        \caption{Comparison of Coverage Area}
        \label{fig: Over protection}
\end{figure}

Currently, database-driven solutions have been used in the U.S. and Europe to manage and allocate spectrum to the SUs while protecting the PUs. These systems collect information on PUs' locations and transmission parameters to define a protection zone based on path-loss models and then facilitate the process of sharing spectrum to SUs.
However, these solutions are inherently conservative and inflexible because (i) they apply overly conservative propagation models across all regions without considering their spatial differences, and (ii) they cannot adjust power allocation to SUs dynamically because of the temporally varying PU operations. As an example, we conducted drive tests in the TVWS band around the Wilson Hall base station. We set up a mobile UE on a truck and measured at different locations, as shown in the top left and right of Fig.~\ref{fig: Over protection}. Database-driven spectrum sharing policy only allows a SU to operate at an EIRP of 16~dBm, resulting in a very small coverage area around the base station, as shown in the bottom left of Fig.~\ref{fig: Over protection}. In comparison, our measurement study shows that the TVWS spectrum is free at the tested region, with a noise level close to -118~dBm. We operate the SU at a higher power level of 42~dBm in accordance with the FCC Experimental License. As a result, the base station is able to cover a much larger area, as shown in the bottom right of Fig.~\ref{fig: Over protection}. Irregularity of the coverage area is due to different terrain 
conditions along different directions.

 Often, overly conservative spectrum models lead to incorrect estimations of available channels. For example, in the ARA footprint, the TVWS database indicates that only three channels are available for fixed wireless deployment, but ground truth measurements show substantial discrepancies, i.e., almost 50\% of spectrum bands are available at almost all locations of extensive ARA footprint spanning over 155 square kilometers. These inaccuracies are the result of overestimated protection zones for PUs, which can limit the efficient use of spectrum by smaller networks similar to those on farms. Additionally, the databases do not adequately capture the dynamic nature of spectrum availability in the temporal domain, further compromising the effectiveness of spectrum sharing and management. 
 
 Furthermore, technical and regulatory hurdles directly contribute to the inadequate protection of SUs from other SUs trying to operate in the nearby geographical areas. Multiple users trying to access the spectrum are often allocated the same spectrum bands leading to harmful interference. One such scenario was reproduced in the 3450--3550 MHz spectrum band on the ARA platform, where we used a Commercial-off-the-Shelf~(COTS) UE as the primary SU and multiple National Instruments (NI) SDRs at nearby farm locations as the secondary SUs. When all the SUs attempt to access the same spectrum band, the throughput on COTS UE was reduced from 598 Mbps to 139 Mbps.
 
On any farmland, different operations or applications may have different network requirements. Allocating the entire available spectrum for operations that only need a fraction of the spectrum, such as IoT deployments on farms, leads to inefficient spectrum use. 
Current database-driven approaches lack the ability to dynamically allocate spectrum to different users with different needs.
More dynamic approach, where users specify their required bandwidth and spectrum managers allocate accordingly, are needed to optimize spectrum use and ensure efficient resource allocation.

\section{{Spatiotemporal Modeling for Spectrum Management}}

To address the limitations of database-driven spectrum sharing mechanisms, spectrum-sensing-based spatiotemporal models are suggested. These models more effectively represent the variations in spectrum availability across both spatial and temporal dimensions~\cite{Radio_Dynamic_Zones}. Integrating advanced modeling techniques into database-driven spectrum management systems enables real-time spectrum adjustments, allowing for optimal spectrum allocation based on actual demand, thus improving spectrum utilization~\cite{UAV_spectrum_map}. Spectrum sensing with limited number of stand-alone cognitive radios faces challenges such as the hidden node problem, which affects the spectrum sensing accuracy. On the other hand, a wide deployment of high-precision RF sensors is often impractical due to the cost constraints and data overhead. Spatiotemporal modeling offers a solution to improve spectrum management with limited number of RF sensors by dynamically mapping the spectrum occupancy across the geographical area of interest.

\subsection{{Existing Spatiotemporal Modeling Techniques}} 
Stochastic models such as Okumura and Hata-Davidson, based on extensive data-driven studies, fail to capture real-world variations impacting signal coverage. Deterministic models, such as ray-tracing, can simulate wave behaviors but require significant computational resources, limiting their use in real-time spectrum sharing in dynamic network settings. Both approaches lack integrating real-world RF sensor data, which is crucial for a comprehensive understanding of spectrum occupancy \cite{PINN}.
Standard Machine Learning (ML) and Deep Learning (DL) techniques are often not generalizable to diverse geographical sites and frequencies, requiring vast amount of labeled data for accurate predictions. Obtaining such datasets is labor-intensive, leading researchers to rely on simulated datasets that fail to capture the real-world complexities effectively. Efficient modeling techniques that can capture spatiotemporal variations with minimal samples are needed.

\begin{figure*}[t!]
        \centering
        \hspace*{-5mm}
        \includegraphics[scale=0.65]{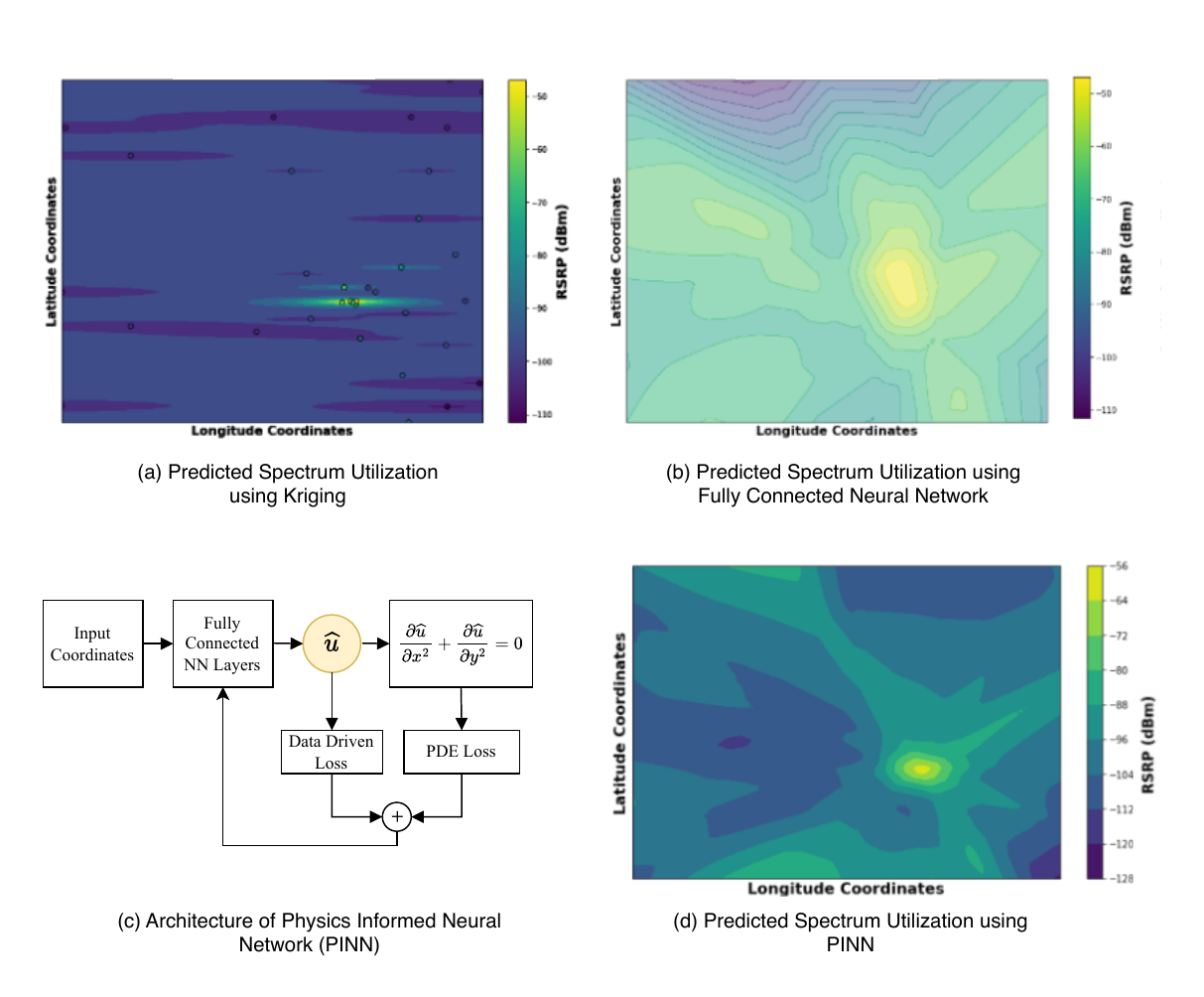}
        \caption{Physics-Informed Neural Network (PINN)}
        \label{fig: PINN}
\end{figure*}

\subsection{{Physics-Informed Neural Network (PINN)}} 

Accurately predicting spectrum usage across large geographical area 
with sparse data is a challenging task. Methods such as Kriging interpolation (which estimates unknown points based on the weighted average of known data) have limitations due to the complexity of data and impractical stationarity assumptions. Tensor decomposition methods also face challenges in modeling non-stationary processes and computational demands. Deep learning can handle non-stationary processes well, but requires extensive dataset. On the other hand, incorporating physical laws into Neural Networks, such as Physics-Informed Neural Networks (PINN) \cite{PINN_Orignal}, enhances accuracy and reduces training time. A PINN embeds physical laws described by Partial Differential Equations (PDEs) into the DL loss function, ensuring the conformance of the network to both data-driven and physical constraints. PDEs such as Maxwell's equations have been used in the literature to model the spatiotemporal variation of electric and magnetic fields in a given region. In our preliminary studies on using PINN for spatiotemporal modeling of wireless spectrum, we use the Laplace equation, a special case of Maxwell's equation at steady state when the time variation is equals to zero. Figs.~\ref{fig: PINN}(a) and ~\ref{fig: PINN}(b) represent the reconstructed radio environment maps (i.e., the spectrum utilization) generated from Kriging and conventional neural network, respectively. The $x$ and $y$ axes represent the longitudinal and latitudinal coordinates, respectively, and the color intensity indicates the expected received signal strength at each location. Fig.~\ref{fig: PINN}(c) shows the architecture of
PINN that takes spatial coordinates as its input. The five fully-connected layers of PINN are trained using both data and PDE driven loss functions. With only 64 training samples, the PINN effectively captures the physical terrain dependencies, as shown in Fig.~\ref{fig: PINN}(d), and outperforms the standard modeling approaches such as Kriging and conventional Neural Networks~(NN). The data samples for PINN training are collected from carefully chosen locations characterized by variations in terrain, foliage, and multipath effect. In short, PINN demonstrates its ability in capturing the complexities of the terrain with the sampled data alone, i.e.,  without any inputs on the geographical information. Additionally, PINN converges at a faster pace with greater accuracy. The testing Mean-Squared-Errors~(MSEs) for Kriging and NN are 97.6 and 14.8, respectively, while PINN has the MSE of only 0.02.

\section{{Spectrum Innovation Opportunities}}

Spectrum sharing based on spatiotemporal spectrum occupancy models can significantly improve the overall spectrum usage efficiency. However, this requires real-time knowledge of the radio environment in the geographical area, facilitated by RF sensors. Unlike controlled environments where signals may be modeled fairly accurately using well-established path loss models, real-world outdoor scenarios are affected by buildings, terrain, 
and environmental factors like weather conditions.

To account for these dynamic variables, new mathematical models using partial-differential-equations ~(PDEs) are needed. These models should incorporate the effects of scattering, reflections, and refraction of signals, in order to enhance the accuracy of PINNs for designing and developing new protocols. Unique measurement data collected from agriculture and rural deployments of ARA can be leveraged to identify optimal differential equations through regression methods such as SINDy~(Sparse Identification of Dynamic Systems)~\cite{SINDy}.

Based on these models, new protocols can be established to allocate spectrum by identifying white spaces in the spatial, temporal, and frequency domains. Users can be matched with spectrum slices that meet their Quality of Service (QoS) and utilization time period requirements. Additionally, spatiotemporal models for spectrum sensing can help dynamically adjust protection zones around users based on real-time RF spectrum usage. This approach minimizes spectral leakage to other users and supports the operations of mobile SUs, which is a challenging task with the current database-driven spectrum-sharing approaches.

\section{{Concluding Remarks}}

Addressing the broadband divide in Rural Farmlands of the U.S$.$ is important for integrating these regions 
to the nation's economic and social fabric. Despite the economic challenges, improving the rural broadband infrastructure for wider coverage that could enable modern agricultural practices and other vital applications for day-to-day life is utterly important. From the measurements, it can be seen that there are abundant spectrum bands that are highly under-utilized and can be used to empower local farmers and communities in operating broadband wireless networks. Dynamic spectrum management through advanced deep learning models such as PINNs can help improve the overall spectrum use efficiency in rural regions.

\bibliographystyle{IEEEtran}
\bibliography{references}

\section*{{Biographies}}

\textbf{Mukaram Shahid} is pursuing his Ph.D. from Iowa State University. He obtained his bachelor's degree in Electrical Engineering from Ghulam Ishaq Khan Institute of Engineering Sciences and Technology, Pakistan, and his master's degree from ISU. His research focuses on practical implications on Dynamic Spectrum Sharing and Spectrum Policy.\\

\textbf{Kunal Das} is pursuing his Ph.D. in Statistics from Iowa State University. He finished his master's from the Indian Statistical Institute. His research interests include non-parametric spatial and spatio-temporal statistics and survey sampling.\\

\textbf{Taimoor Ul Islam} is pursuing his Ph.D. from Iowa State University in computing and networking systems. He obtained his bachelor's and master's degree in Electrical Engineering from National University of Sciences and Technology (NUST), Pakistan and his current research explores practical problems in mMIMO systems.
\\

\textbf{Christ Somiah} completed his bachelor's in Telecommunications Engineering from Kwame Nkrumah University and pursuing his master's from Iowa State University. He focuses on spectrum sharing and wireless security.
\\

\textbf{Daji Qiao} is a Professor in the Department Electrical and Computer Engineering at Iowa State University. He obtained his Ph.D. from the University of Michigan.  His research focuses on computing and networking systems, wireless networking, and cyber security.
\\

\textbf{Jiming Song} is a Professor in Electrical and Computer Engineering at Iowa State University. He obtained his Ph.D. from Michigan State University. His research focuses on electromagnetic and ultrasonic  wave propagations and non-destructive evaluation.
\\

\textbf{Arsalan Ahmad} is a Research Associate Professor in Electrical and Computer Engineering at Iowa State University. He earned his Ph.D. from Politecnico di Torino, Italy with specialization in computer networking.
\\

\textbf{Yong Guan} is a Professor in Electrical and Computer Engineering at Iowa State University and Fellow of AAFS. He obtained his Ph.D. from Texas A\&M University and his research includes digital forensics, wireless security, and cyber law.
\\

\textbf{Zhengyuan Zhu} is a Professor in Statistics at Iowa State University. He received his Ph.D. from the University of Chicago and currently focuses on spatial statistics, survey statistics, and machine learning.
\\

\textbf{Tusher Chakraborty} is a Research Software Engineer at Microsoft Research in Redmond, USA focusing on real-world networking systems.
\\

\textbf{Suraj Jog} is a Senior Researcher at Microsoft Research in Redmond, USA focusing on 6G and space networking. He earned his Ph.D. from University of Illinois Urbana-Champaign.
\\

\textbf{Sarath Babu} currently working as a Research Scientist at Iowa State University focusing on the design and deployment of next generational networking infrastructures. He obtained his Ph.D. from the Indian Institute of Space Science and Technology.
\\

\textbf{Ranveer Chandra} currently serving as the Managing Director for Research for Industry, and the CTO of Agri-Food at Microsoft. He earned his Ph.D. from Cornell University, USA.
\\

\textbf{Hongwei Zhang} is the Director of the Center for Wireless, Communities and Innovation (WiCI) and serves as the Willard and Leitha Richardson Professor at Iowa State University. He earned his Ph.D. from The Ohio State University, USA.

\end{document}